\documentclass[aps,prl,reprint, preprintnumbers,amsmath,amssymb,superscriptaddress,floatfix,showpacs]{revtex4-1}

\usepackage{graphicx}
\usepackage{dcolumn}
\usepackage{bm}
\usepackage{hyperref}
\usepackage[utf8x]{inputenc} 
\usepackage{textcomp} 
\usepackage{comment} 
\usepackage{bookmark}

\newcommand{\ex}{{\bf e}_x}
\newcommand{\ey}{{\bf e}_y}
\newcommand{\ez}{{\bf e}_z}

\usepackage{xcolor}

\def\gamdinf{{\dot{\gamma}_\infty}}

\begin{document}


\title{Resonant alignment of microswimmer trajectories in oscillatory shear flows}

\author{Alexander Hope}%
\affiliation{Department of Chemical and Process Engineering, University of Strathclyde,
James Weir Building, 75 Montrose Street, Glasgow G1 1XJ, United Kingdom}
\author{Ottavio A. Croze}
\email{oac24@cam.ac.uk}
\affiliation{Cavendish Laboratory, University of Cambridge, Cambridge, CB3 0HE, United Kingdom}
\author{Wilson C. K. Poon}
\affiliation{SUPA and The School of Physics \& Astronomy, The University of Edinburgh,
Kings Buildings, Mayfield Road, Edinburgh EH9 3JZ, United Kingdom}
\author{Martin A. Bees}
\affiliation{Department of Mathematics, University of York, York YO10 5DD, United Kingdom}
 \author{Mark D. Haw}
\affiliation{Department of Chemical and Process Engineering, University of Strathclyde,
James Weir Building, 75 Montrose Street, Glasgow G1 1XJ, United Kingdom}


\date{\today} 

\begin{abstract}
Oscillatory flows are commonly experienced by swimming microorganisms in the environment, industrial applications and rheological investigations. We experimentally characterise the response of the alga {\it Dunaliella salina} to oscillatory shear flows, and report the surprising discovery that algal swimming trajectories orient perpendicular to the flow-shear plane. The ordering has the characteristics of a resonance in the driving parameter space. 
The behaviour is qualitatively reproduced by a simple model and simulations accounting for helical swimming, providing the mechanism for ordering and criteria for the resonant amplitude and frequency. The implications of this work for active oscillatory rheology and industrial algal processing are discussed.

\end{abstract}
 
\pacs{Valid PACS appear here}

\maketitle

Many swimming microorganisms experience shear flow in natural and industrial processes. Swimming is strongly biased by environmental cues and fluid shear \cite{PedleyKessler92, RusconiStocker15}, with significant implications for ecology \cite{Durhametal09} and industrial exploitation \cite{BeesCroze14}. Classic examples include directed swimming relative to light (phototaxis) and hydrodynamic focusing in down-welling flow due to viscous and gravitational torques (gyrotaxis) \cite{PedleyKessler92,HillPedley05}. 

There is great potential to use individual and collective microswimmer behaviour to improve microbial biotechnology, such as algal photobioreactor design \cite{BeesCroze14}. For example, gyrotactic microorganisms in laminar channel flow tend to focus and so drift faster and diffuse less than non-swimming cells or nutrients \cite{BearonBeesCroze12, Crozeetal13} while in turbulent flows cells accumulate in transiently downwelling \cite{Crozeetal13, Durhametal13, Zhanetal14} or strongly accelerated \cite{DeLilloetal14} regions.  Horizontal shear flows can trap gyrotactic swimmers (a mechanism for oceanic thin layers) \cite{Durhametal09} and modify hydrodynamic instabilites and patterns (bioconvection) \cite{CrozeAshrafBees10, HuangPedley14}.  Phototaxis and shear flow can combine to drive cell focusing \cite{Garciaetal13} and pattern formation \cite{WilliamsBees11}. Complex transport dynamics can even result from relatively simple shear flow \cite{ZoettlStark12, RusconiStocker15}. The rheology of active media is also of recent interest: suspensions of swimming bacteria behave less viscously \cite{SokolovAranson09} and algae more viscously \cite{RafaiJibutiPeyla10} than dead cells. 

Here, we investigate the interaction of the swimming alga {\it Dunaliella salina} with oscillatory shear flows. Surprisingly, in experiments swimming trajectories are strongly ordered by the flow for particular driving parameter values. The ordering mechanism is distinct from that observed recently with {\it Dunaliella primolecta}, with constant, strong shear flows \cite{Chengalaetal13}.  We explore the observed resonant ordering employing simple but predictive models, and discuss implications for active oscillatory rheology and industrial processing of swimming algae. 
%
\begin{figure}
\includegraphics[width = 0.33\columnwidth, angle =90]{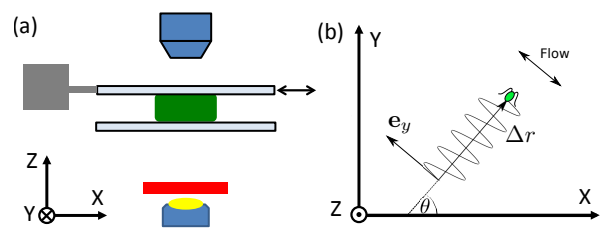}
\caption{\label{Fig1_expsetup} (a) Schematic of apparatus to image suspensions of algae in oscillatory shear flows. (b) Flow oscillation axis $\ey$ (not always aligned with the image $x$ axis, but consistent across repeat experiments) and the angle $\theta$ of the swimming trajectory displacement $\Delta\mathbf{r}$ to the $x$-axis.}
\end{figure}

{\it D. salina} (CCAP 19/18) cells were grown on modified Pick medium \cite{Picketal86, PollePriv} under $12$:$12$ light/dark cycle at 21$^\circ$C. All experiments were carried out at this temperature. Cells were concentrated by upswimming (gravitaxis) using cotton wool rafts \cite{CrozeAshrafBees10}. 
Dilute ($10^6$ cells/ml) suspensions were subjected to oscillatory shear on the stage of an optical microscope (Olympus BX51). The suspension was placed between two transparent parallel plates 400 $\mu$m apart, the top plate connected to an electromechanical drive that sinusoidally sheared the suspension (Figure \ref{Fig1_expsetup}). Plate parallelism was ensured by zeroing sample capillary flow. 
Video sequences of sheared algae were acquired using a Mikrotron MC1310 at $10\times$ (NA 0.25) using red-filtered bright-field illumination to minimise phototaxis \cite{Martinezetal12}. Sequences were captured in a plane equidistant from top and bottom plates at depth 200 $\mu$m. Algae were tracked using MATLAB versions of established algorithms \cite{Kilfoil, CrockerGrier96}. The direction of the imposed oscillating flow was inferred from short-time cell trajectories and confirmed by tracking PEGylated polystyrene colloids (Supplementary Materials, FigS1). Cell observation before and after measurements found that the apparatus did not damage the cells (e.g.~deflagellation).

\begin{figure}
\includegraphics[width = \columnwidth]{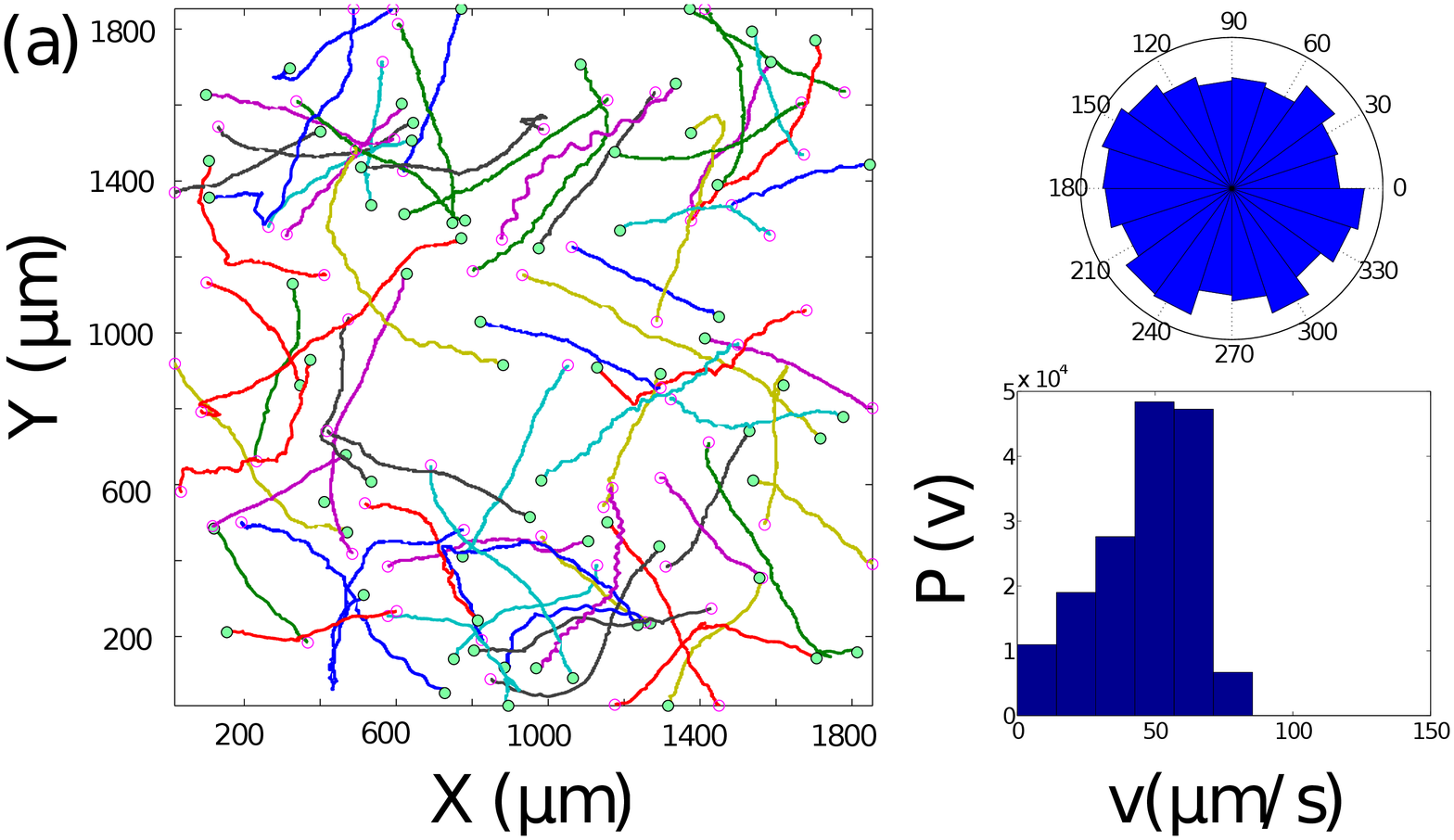} 
\includegraphics[width = \columnwidth]{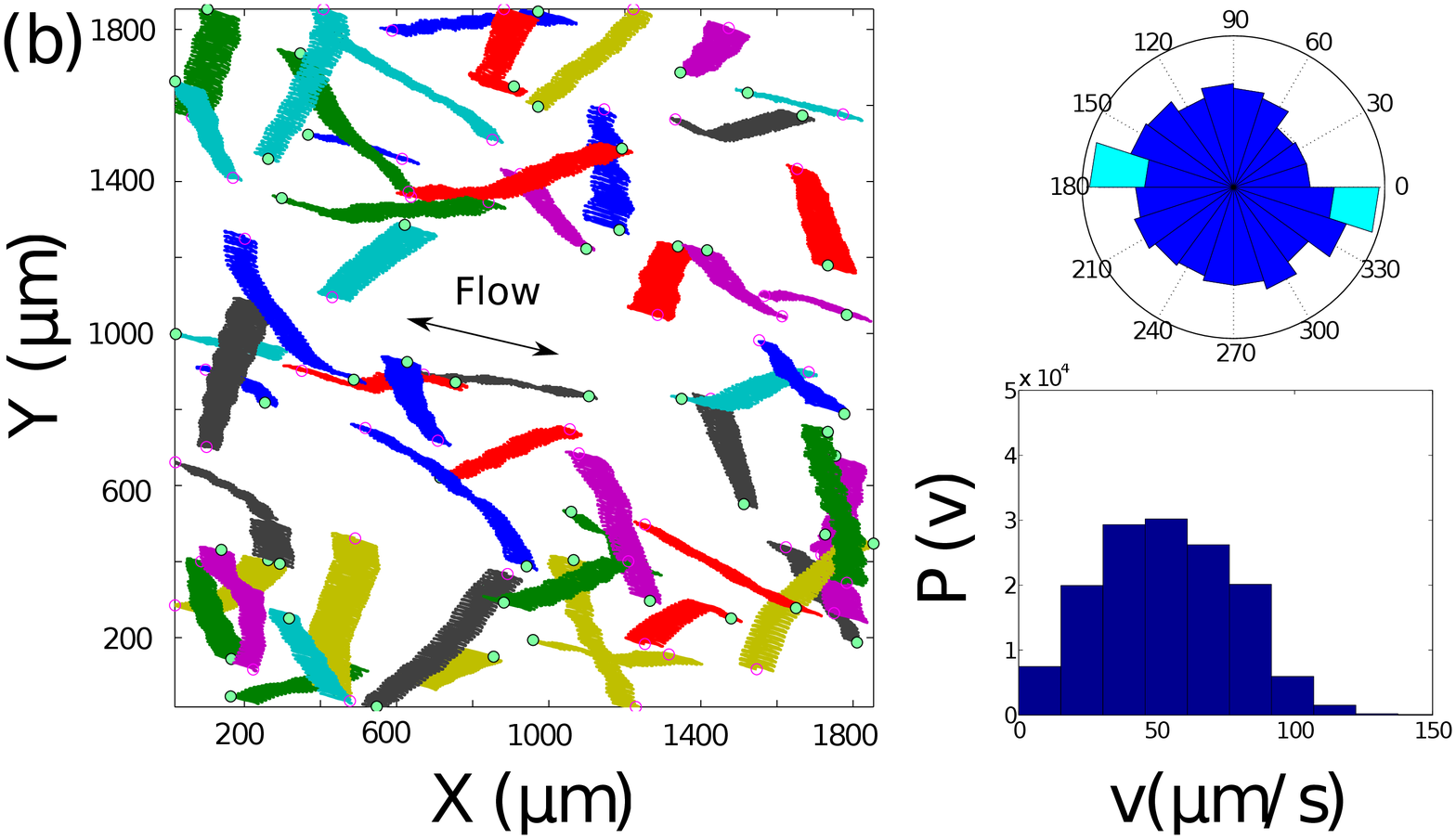}
\includegraphics[width = \columnwidth]{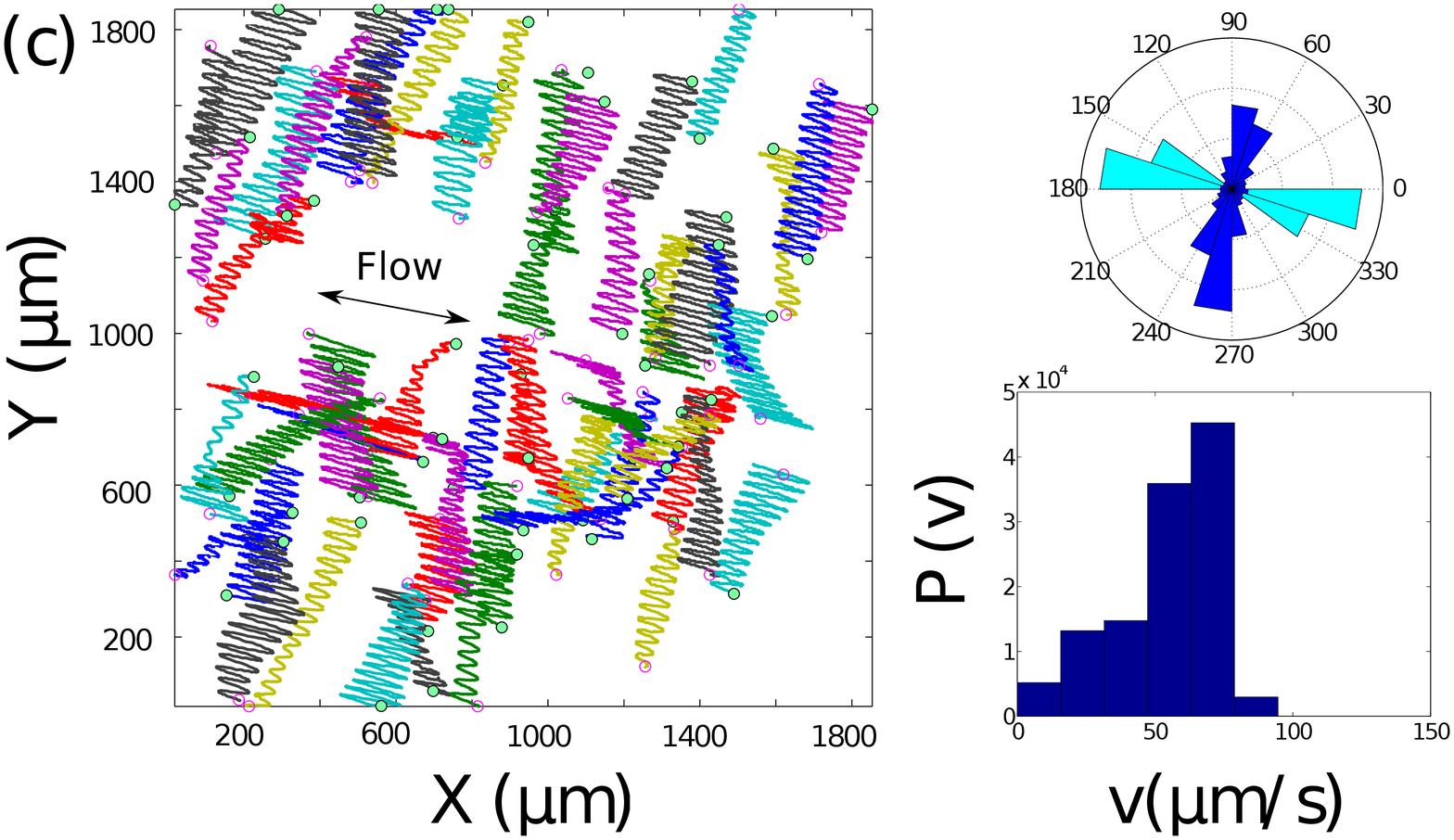}
\caption{\label{Fig2_tracks} Trajectories, orientation distributions and speed of {\it D. salina}. (a) No flow: isotropic trajectories. (b) Oscillatory shear flow with amplitude $A=120$ $\mu$m and frequency $f=6$ Hz: tracks show oscillation at short times, but retain isotropy of swimming orientation over longer times. (c) $A=224$ $\mu$m and $f=2$ Hz: swimming directions align perpendicular (blue) to the flow oscillation direction (cyan). Pink/green circles denote start/end points of each $5$s track.}
\end{figure}

With no flow, {\it D. salina} swimming trajectories were distributed isotropically in the horizontal plane (Figure \ref{Fig2_tracks}a). As in \cite{HillHaeder97}, gravitactic bias was not evident for tracks in this plane on experimental timescales. On application of oscillatory shear with amplitude $A$ and frequency $f$ trajectories might be expected simply to reflect superposition of isotropic swimming and oscillatory advection.  This was indeed the case for some driving parameters, such as $A=120$ $\mu$m and $f=6$ Hz (Fig \ref{Fig2_tracks}b). However, for $A=224$ $\mu$m and $f=2$ Hz swimming trajectories unexpectedly aligned perpendicular to the flow-shear plane (Fig \ref{Fig2_tracks}c).
%
%
Alignment can be quantified by the start-to-end displacement vector, $\Delta \mathbf{r}(\tau)$, of the trajectory of each swimmer (see \ref{Fig1_expsetup}b). $\Delta \mathbf{r}(\tau)$ evaluated at short times within oscillation cycles provides the oscillating flow direction $\ey\sim\Delta \mathbf{r}(\tau\to0)$. However, evaluating $\Delta \mathbf{r}(\tau)$ over the largest available integer multiple $n$ of the oscillation period, $\tau=n/f$, excludes cycle-by-cycle flow oscillations and provides the orientation $\theta$ of the swimming trajectory; i.e. it measures net swimming progress. 

Distributions $P(\theta)$ of orientations are presented beside the trajectories in Figure \ref{Fig2_tracks}. The distribution for $f=2$ Hz, $A=224$ $\mu$m shows how trajectories orient along the line perpendicular to the flow-shear plane, but are equally likely in either direction along this line. If the flow is halted, the distribution returns to uniform (Figure S1, Supplementary Materials). Swimming speed distributions, also shown in Figure \ref{Fig2_tracks}, appear to accentuate the importance of strong swimmers during alignment. However, for all cases modal speed is $\approx$60 $\mu$m s$^{-1}$, which compares well with recent measurements \cite{DDMhelicalinprep}.



\begin{figure}
\includegraphics[width = \columnwidth]{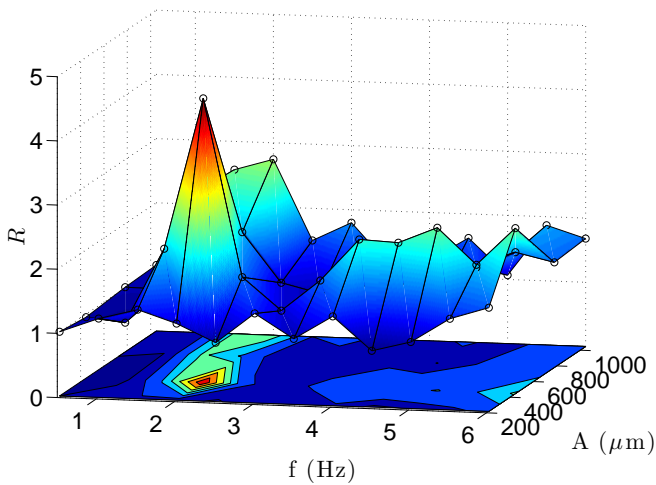}\\
\caption{\label{Fig3_ordershear} Alignment ratio $R$ as a function of driving amplitude $A$ and frequency $f$ for oscillatory flow in a shear cell with $H=400$ $\mu$m gap width. $A$ is a measure of horizontal displacement.}
\end{figure}

\begin{figure}
\includegraphics[width =\columnwidth]{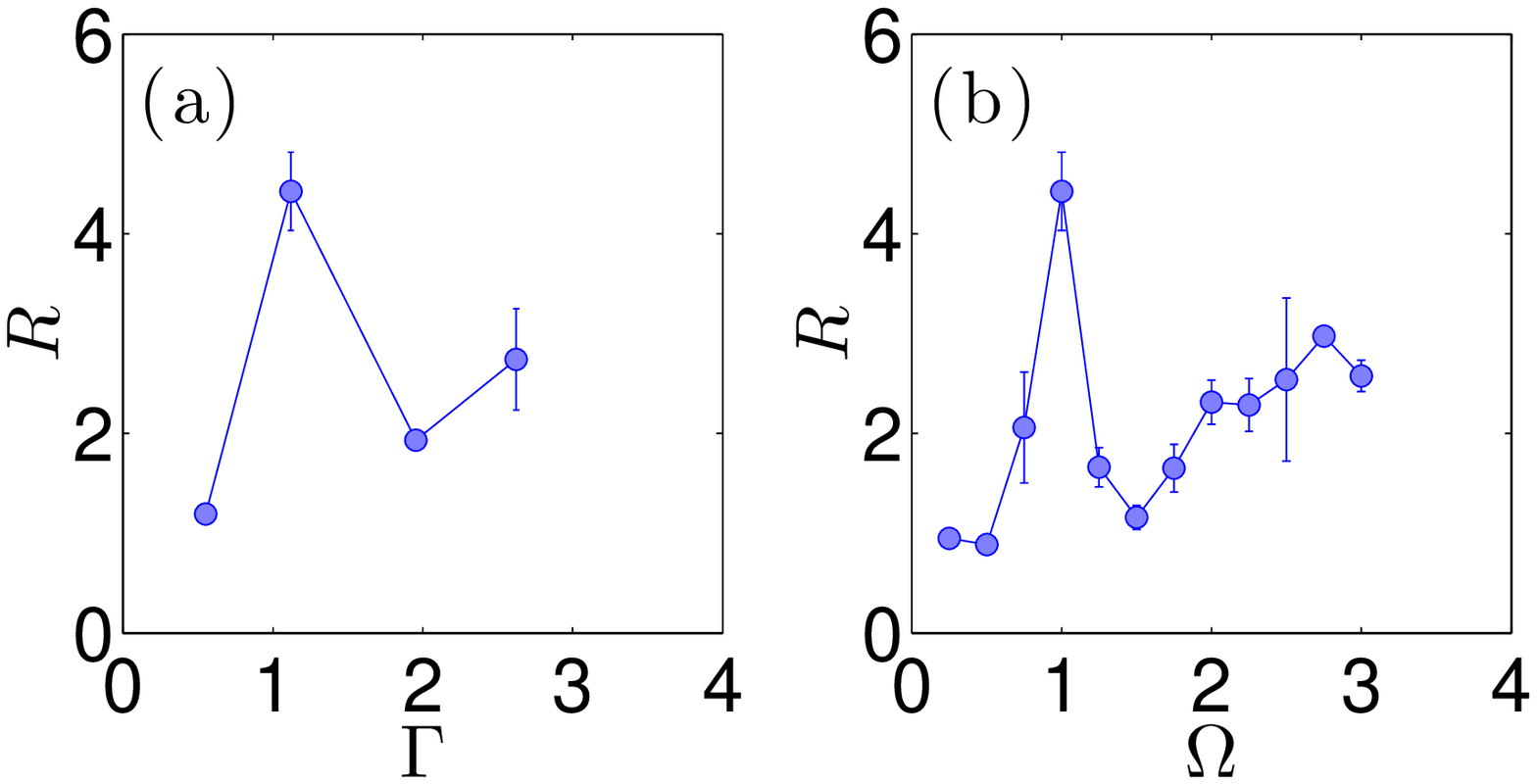}\\
\includegraphics[width =\columnwidth]{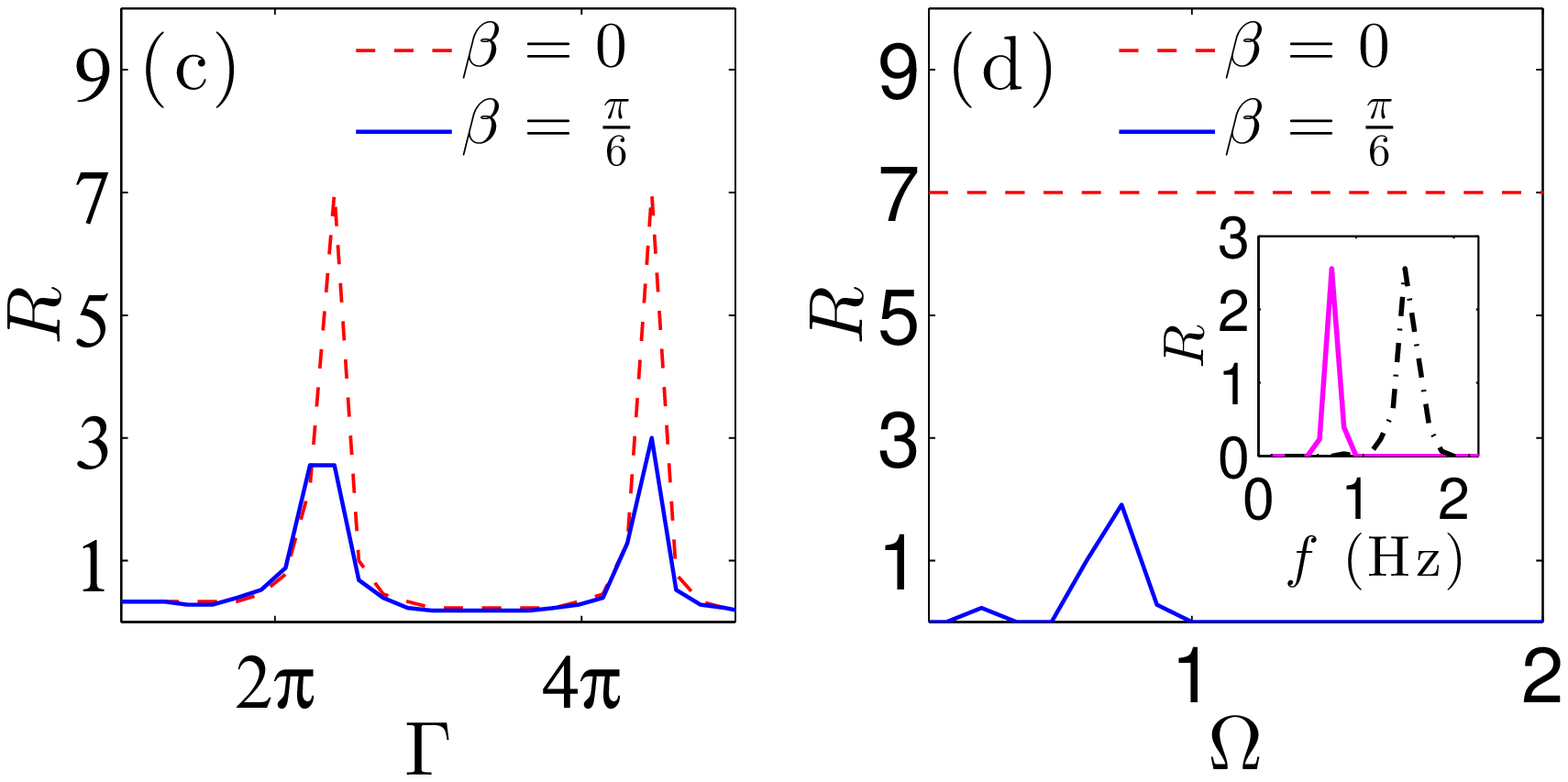}\\
\includegraphics[width =\columnwidth]{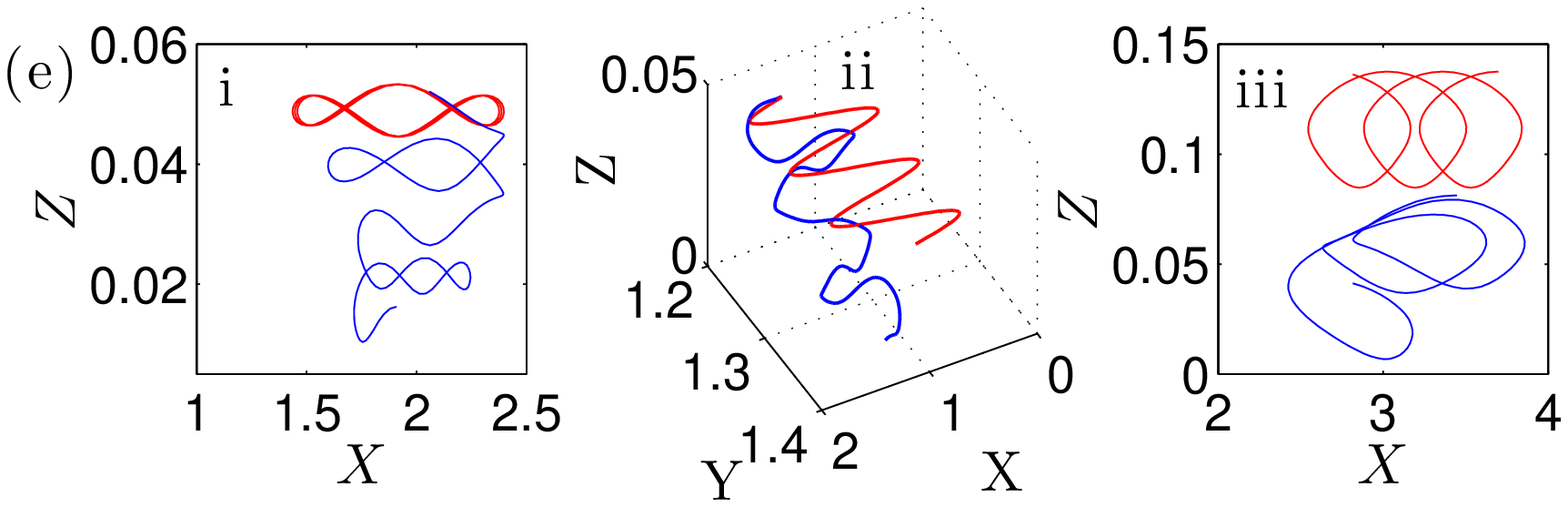}
\caption{\label{Fig4_simulations} 
Alignment ratio $R$ as a function of amplitude $\Gamma$ and frequency $\Omega$ from experiment (a, b) and simulation (c, d). Simulation predictions are shown with ($\beta=\pi/6$, solid) and without ($\beta=0$, dashed) helical swimming (frequency $f_h=2$Hz), only the latter providing frequency dependence. Trajectories are also shown in (e): at resonance, $[\Gamma, \Omega]=[7.5,0.75]$, for $XZ$ plane (i) and 3D views (ii); off-resonance $[\Gamma, \Omega]=[4,0.75]$ (iii). Only trajectories closed in $XZ$ provide ordering. Inset of (d): simulations of {\it Pfiesteria piscida} in oscillatory flow. When predating, this dinoflagellate increases its helical frequency from $1$ (solid line) to $2$ Hz (dash dot) \cite{Shengetal07}. We predict a measureable shift in the resonance peak.}
\end{figure}

To statistically quantify observed alignment as a function of the flow parameters, we count trajectories with displacement $\Delta \mathbf{r}$ oriented perpendicular, $N_\perp$, and parallel, $N_\parallel$, to the flow direction and define $R=N_\perp/N_\parallel$. By parallel (perpendicular) displacements we mean swimming orientations within $\pm \pi/4$ of the flow (vorticity) axis. The surface plot in Figure \ref{Fig3_ordershear} illustrates how the alignment has the characteristics of resonance, occupying a small region of the imposed flow parameter space. Fixed amplitude and frequency sections of the ordering surface are shown in Figure \ref{Fig4_simulations} and discussed below.


%
%

{\it Discussion.} The observed trajectories result from the combination of shear flow and swimming. 
Many algae swim helically, in part to facilitate phototaxis via a directional eyespot \citep{FosterSmyth80}. Thus we model the algae as helical swimmers in a flow. Following \citep{Bearon13}, we assume that a cell at position $\mathbf{r}$ swimming with speed $v$ in direction $\mathbf{p}$ has an intrinsic angular velocity $\omega_h \mathbf{n}$ about an axis $\mathbf{n}$, where $\mathbf{p}\cdot \mathbf{n}=\cos{\beta}$, for constant angle $\beta$, due to an asymmetric (non-planar) flagellar stroke.  Hence, if the only external torque is due to flow with velocity $\mathbf{u}$ and vorticity $\boldsymbol\omega$, spherical swimmers obey $\dot {\mathbf{r}}= \mathbf{u}(t) + v \mathbf{p}$, $\dot{\mathbf{p}} = \left(\frac{1}{2}\boldsymbol\omega(t)+\omega_h\mathbf{n} \right)\times \mathbf{p}$ and $\dot{\mathbf{n}}  = \frac{1}{2}\boldsymbol\omega(t) \times \mathbf{n}$. Experimental observations suggest that the fluid velocity is $\mathbf{u}(t)=\gamdinf Z \cos (\omega_d t) \ex$, with vorticity $\boldsymbol\omega(t)=\gamdinf \cos (\omega_d t) \ey$. Here $\ex$ and $\ey$ are along the positive flow and vorticity directions, respectively, $\gamdinf=\omega_d A/H$ is the maximum shear rate and $\omega_d=2\pi f$ is the angular driving frequency (recall $A$, $f$ and $H$ are driving amplitude, frequency and gap width, respectively). Nondimensionalising lengths with $A$ and times with $1/\omega_d$, the model equations read
\begin{eqnarray}
&&\dot {\mathbf{r}}= \Gamma \cos (t) Z\ex +\nu \mathbf{p}\label{eq:posvect}\\
&&\dot{\mathbf{p}} = \left[\frac{\Gamma}{2}  \cos (t) \ey+\Omega^{-1}\mathbf{n} \right]\times \mathbf{p};  \mbox{~}\dot{\mathbf{n}}  = \frac{\Gamma}{2}  \cos (t)\ey \times \mathbf{n} \mbox{~~~} \label{eq:orientvect}
\end{eqnarray}
where $\Gamma=\gamdinf/\omega_d=A/H$ is the dimensionless shear rate/amplitude,  $\Omega=\omega_d/\omega_h$ is a frequency ratio and $\nu=v/(A \omega_d)=\nu_0/(\Gamma \Omega)$ a nondimensional swimming speed, with $\nu_0=v/(\omega_h H)$. The Cartesian representation of (\ref{eq:posvect}-\ref{eq:orientvect}), with $Z$ measured from the bottom plate and $XY$ the flow-vorticity plane (figure \ref{Fig1_expsetup}), was solved numerically to simulate a suspension of swimmers (see Supplementary Materials). The ratio $R$ was computed from many simulations with a uniform distribution of initial swimmer orientations. 

Non-swimmers ($\nu\to0$) passively follow the imposed oscillatory flow. In the absence of flow, the governing equations predict helical trajectories: $\mathbf{p}$ rotates around $\mathbf{n}$ (the helix axis) with frequency $\omega_h$.  With the experimental resonance close to  helical swimming frequency, it is tempting to think that helical trajectories are responsible for the observed alignment. This is only partially true, as we shall see. 

First we ask if the combination of `non-helical' swimming and oscillatory shear alone is sufficient to induce the observed behaviour. Figure \ref{Fig4_simulations} displays the ordering ratio (experiment a,b; simulation c,d) and simulated swimmer trajectories (e) for a model where helical swimming is switched off (dashed lines). While $R$ peaks at characteristic values of the driving amplitude $\Gamma$ (Fig \ref{Fig4_simulations}c), it is entirely independent of driving frequency (Fig \ref{Fig4_simulations}d) in stark contrast to the experiments. The non-helical limit does, however, predict alignment: it is worth considering further.
In this limit ($\beta\to0$), the governing equations simplify considerably if we choose Euler angles $\Theta$ and $\Phi$ such that $\mathbf{p}=\mathbf{n}=(\sin\Theta\sin\Phi, \cos\Theta, \sin\Theta \cos\Phi)$, where $\Theta$ increases from the direction of vorticity, $\ey$, along the $Y$-axis, and $\Phi$ is measured from the $Z$-axis [e.g. ($\Theta$, $\Phi$)= ($\pi/2$, $\pi/2$) is along the $X$-axis]. Equations (\ref{eq:posvect}-\ref{eq:orientvect}) give
\begin{eqnarray} 
\dot{X}&=& \nu_\perp \sin\Phi +  \Gamma \cos (t) Z \label{eq:jefftrajX}\\
\dot{Y}&=& \nu_\parallel  \label{eq:jefftrajY}\\
\dot{Z}&=& \nu_\perp \cos\Phi \label{eq:jefftrajZ}\\
\dot{\Theta}&=&0;\,\, \dot{\Phi}=\frac{\Gamma}{2} \cos (t) \label{eq:jefftrajphi}
\end{eqnarray}
where $\nu_\perp=\nu\sin\Theta_0$ and $\nu_\parallel=\nu\cos\Theta_0$ are the nondimensional swimming speed components perpendicular and parallel to $\ey$. Integration of  (\ref{eq:jefftrajY}) and (\ref{eq:jefftrajphi}) yields $Y(t)=Y_0+\nu_\parallel t$, $\Theta(t)=\Theta_0$ and
\begin{equation}\label{eq:phiorder}
\Phi(t)=\Phi_0+ \frac{\Gamma}{2} \sin (t).
\end{equation}
(Recall that $\Gamma=A/H$ is the non-dimensional shear rate.) 
As the $Y$ component of the trajectory grows linearly in time, independent of shear, alignment can only depend on the coupled $X$ and $Z$ dynamics. In particular, closed orbits in the $XZ$-plane are present at resonance, see figure \ref{Fig4_simulations}e, panel (i). For such orbits, progress only in the $Y$-direction is possible, leading to alignment. Off-resonance, orbits are open and cells can progress in $X$ and $Z$ directions (figure \ref{Fig4_simulations}e,  panel (ii)). 

This phenomenology can be understood in terms of oscillatory Jeffery dynamics of the swimmer orientation.
We see from (\ref{eq:phiorder}) that oscillatory shear forces swimmer orientation in the vertical $XZ$-plane to describe circular arcs swept sinusoidally in time 
(contrast this with circular Jeffery orbits in steady shear flow, $\omega_d\to\infty$), with angular amplitude $\Gamma/2$. Folded orbits only arise when shear is sufficiently large to rotate swimmer orientation by integral multiples of $\pi$, so it can make no net progress during a cycle (see \ref{Fig4_simulations}e). This provides a prediction for the resonant ordering amplitude 
$\Gamma_{\rm res} \approx 2 \pi n$, $n\in \mathbb{Z}$, in good agreement with non-helical simulations (Fig \ref{Fig4_simulations}c). The latter agree qualitatively with the experimental results in Fig \ref{Fig4_simulations}a. Quantitatively, smaller values of $\Gamma$ are sufficient to induce ordering in experiment. This could be due to second order mechanisms neglected in our simple model, as discussed below.

The model for non-helical swimmers predicts alignment as a function of amplitude, but does not reproduce the experimentally observed dependence on both driving amplitude {\it and} frequency (Fig \ref{Fig4_simulations}b). {\it D. salina} is a helical swimmer, rotating at $1.5-2$ Hz \cite{DDMhelicalinprep}. This second, internal frequency provides the possibility of further resonance. Indeed, with 
$\beta \neq 0$ (recall $\beta$ is the angle between $\mathbf{p}$ and $\mathbf{n}$)
numerical results reveal that ordering is frequency dependent (Figure \ref{Fig4_simulations}d).
Position equations are unchanged, but depend on 
more complex
orientation dynamics resulting from the coupling of flow-induced and helical rotation. Cell orientation angles $\Theta_p$ and $\Phi_p$, defining $\mathbf{p}$, evolve according to (see Supplementary Materials) 
%
\begin{eqnarray} 
\dot{\Theta}_p&=&\Omega^{-1}\,\sin\Theta_n \sin(\Phi_n-\Phi_p),\label{eq:heltheta}\\ 
\dot{\Phi}_p&=&\frac{\Gamma}{2} \cos (t) \label{eq:helphi} \\ \nonumber 
&+&\Omega^{-1}\,[\cos\Theta_n-\sin\Theta_n\cot\Theta_p\sin(\Phi_n+\Phi_p)], 
\end{eqnarray}
whereas angles for $\mathbf{n}$ satisfy (\ref{eq:jefftrajphi}), such that $\dot{\Theta}_n=0$, $\dot{\Phi}_n=\Gamma \cos (t)/2$. 
As the $\mathbf{p}$-dynamics are slaved to the $\mathbf{n}$-dynamics, 
trajectories with helical swimming do retain broad features of non-helical orbits in oscillatory shear, see Figure \ref{Fig4_simulations}e, 
but they are nevertheless qualitatively perturbed (even for infinitesimal $\beta$). Thus only for particular driving frequencies and amplitudes does helical swimming produce alignment-inducing orbits. The frequency condition for ordering can be obtained by considering the case of a swimmer with $\mathbf{n}$ in the direction of vorticity: $\Theta_n(0)=0=\Phi_n(0)$; $\Theta_p(0)=\beta$; and $\Phi_p(0)=0$. Equation (\ref{eq:helphi}) then integrates to $\Phi_p(t)= \frac{\Gamma}{2} \sin (t) + \Omega^{-1}t$:
a helical phase can perturb simple Jeffery rotation by the flow. Only when $\Omega\sim 1$, i.e. when the driving and helical phase are synchronised, can a resonant value of $\Gamma$ give alignment, agreeing with both simulations and experiments (Fig \ref{Fig4_simulations}b,d). 

Oscillatory Jeffery orbits and helical swimming provide a first order explanation for the dependence of resonance in experiments on both frequency and amplitude. 
The model is predictive: swimmers with different helical frequencies should display a different resonance spectrum (Fig \ref{Fig4_simulations}d, inset). 
This description could be extended to investigate how additional effects such as taxes, orientation noise, inertia and cell shape either dominate, compete or act in concert with helical swimming to affect resonance.  Intriguing possibilities include stochastic resonance due to noise in the flow velocity gradients \cite{GuzmanLastraSoto12}, the active response of flagella to shear \cite{RusconiStocker15, Jibutiteal14}, and effective shape and response to shear due to flagellar beats \cite{OMalleyBees12}.  A full analytical investigation of the nonlinear dynamics of the helical model in oscillatory shear is beyond the scope of this paper, but it is clear that much is to be discovered, analogous to structures observed for swimming cells in Poiseuille flow \citep{ZoettlStark12}.

{\it Conclusions.} We have demonstrated the surprising response of swimming microalgae to oscillatory shear flows, producing an alignment of trajectories with a set of resonance peaks in the parameter space of driving frequency and amplitude. 
A simple model combining shear and non-helical swimming predicts resonant alignment of the trajectories of swimming cells, but only when helical swimming is included does the experimentally observed frequency dependence of the resonance arise. The rich dynamics  and the counterintuitive interactions between swimmers and flow revealed by our experiments and modelling have implications for both active suspension rheology and the design of novel cell processing methods. While simplified models of swimmers (rod or spheroidal pushers and pullers with no helical motion) appear adequate to explain active rheological phenomena such as shear-thickening in algal suspensions \cite{RafaiJibutiPeyla10, Marchettietalreview15}, the current work suggests such models may fail in active {\it oscillatory} rheology experiments. Biotechnologically, the results hint at methods for improvement in efficiency of the algal processing pipeline \cite{ScottetalBiodieselReview10}. For example, in downstream processing of useful microalgae, like the $\beta$-carotene producer {\it D. salina}, cells commonly experience oscillatory and squeezing flows \cite{SqueezingNote}. Resonant alignment will provide boundary accumulation over times $~L/v$, where $L$ is the size of the shear plate and $v$ is the swimming speed. This may be `engineered out' by tuning process parameters from resonance; or it may be fruitfully exploited as a new way to guide and harvest cells.
\begin{acknowledgments}
We thank A. Schofield for providing colloids. OAC, WCKP, MDH and MAB acknowledge support from the Carnegie Trust for the Universities of Scotland, the Winton Programme for the Physics of Sustainability (OAC) and Leverhulme Trust (MDH). OAC and MAB acknowledge an EPSRC mobility grant (EP/J004847/1), WCKP the Programme Grant (EP/J007404/1) and ERC Advanced Grant (ADG-PHYAPS).
\end{acknowledgments}


\begin{thebibliography}{34}
\bibitem [{Squ()}]{SqueezingNote}%
  \BibitemOpen
  \href@noop {} {}\bibinfo {note} {Preliminary squeezing results confirms our
  findings for shear}\BibitemShut {NoStop}%
\end{thebibliography}

\providecommand{\noopsort}[1]{}\providecommand{\singleletter}[1]{#1}%

\newpage

\appendix*

\section{SUPPLEMENTARY MATERIALS}

\section{Shear effects on swimming and flow}

\begin{figure*}
(a) \includegraphics[width = 0.47\columnwidth]{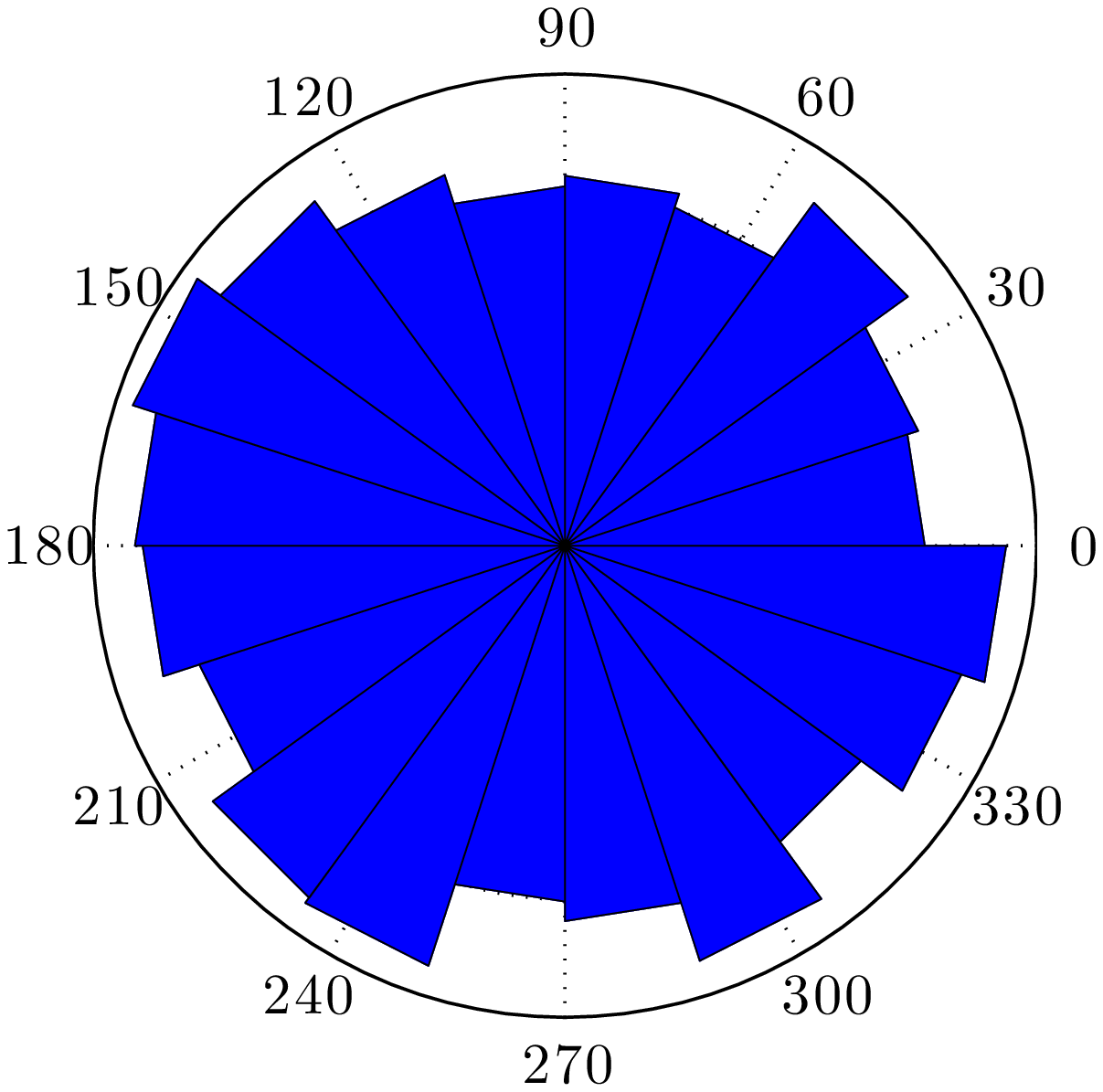} 
\includegraphics[width = 0.47\columnwidth]{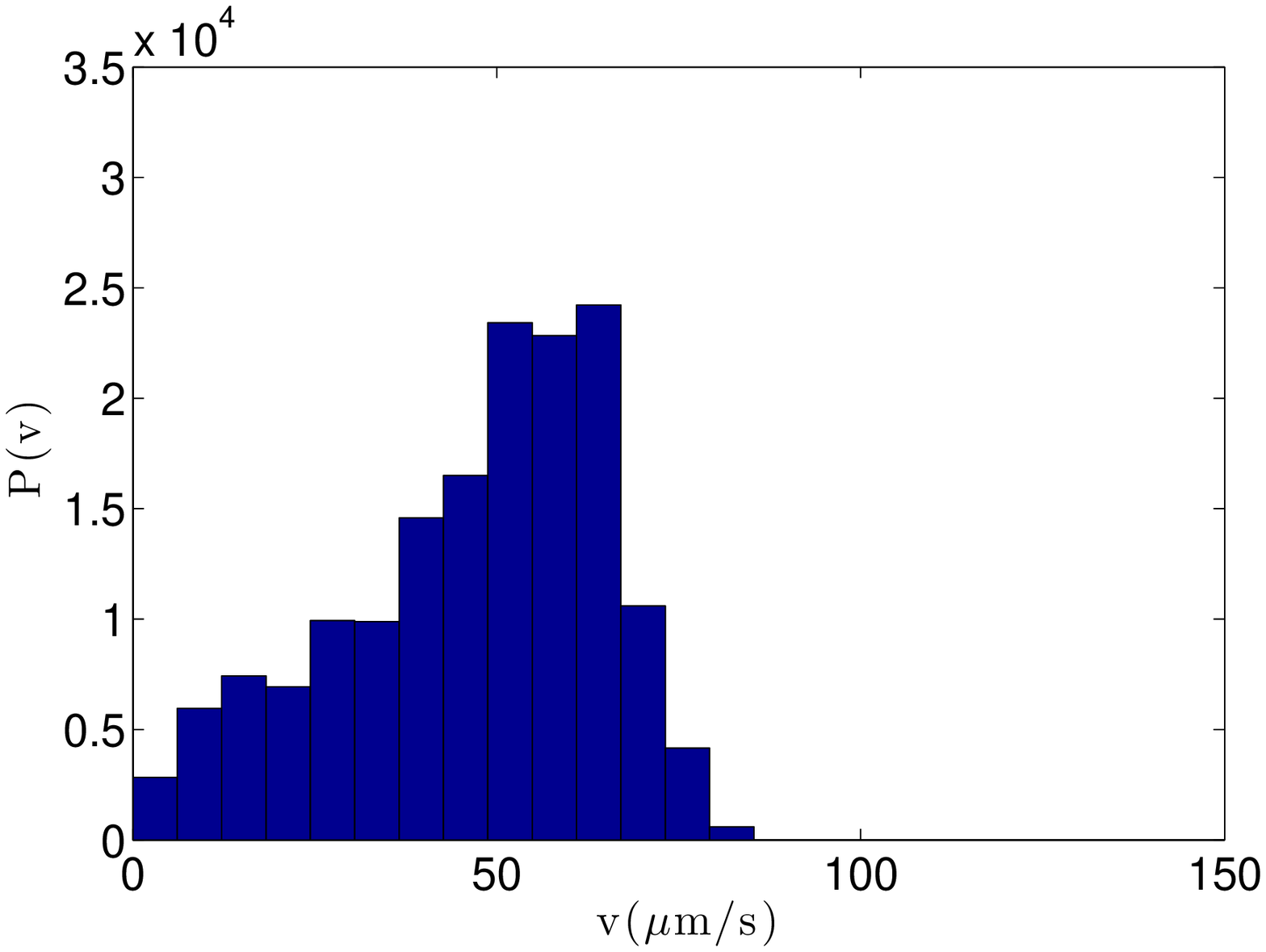}
(b) \includegraphics[width = 0.47\columnwidth]{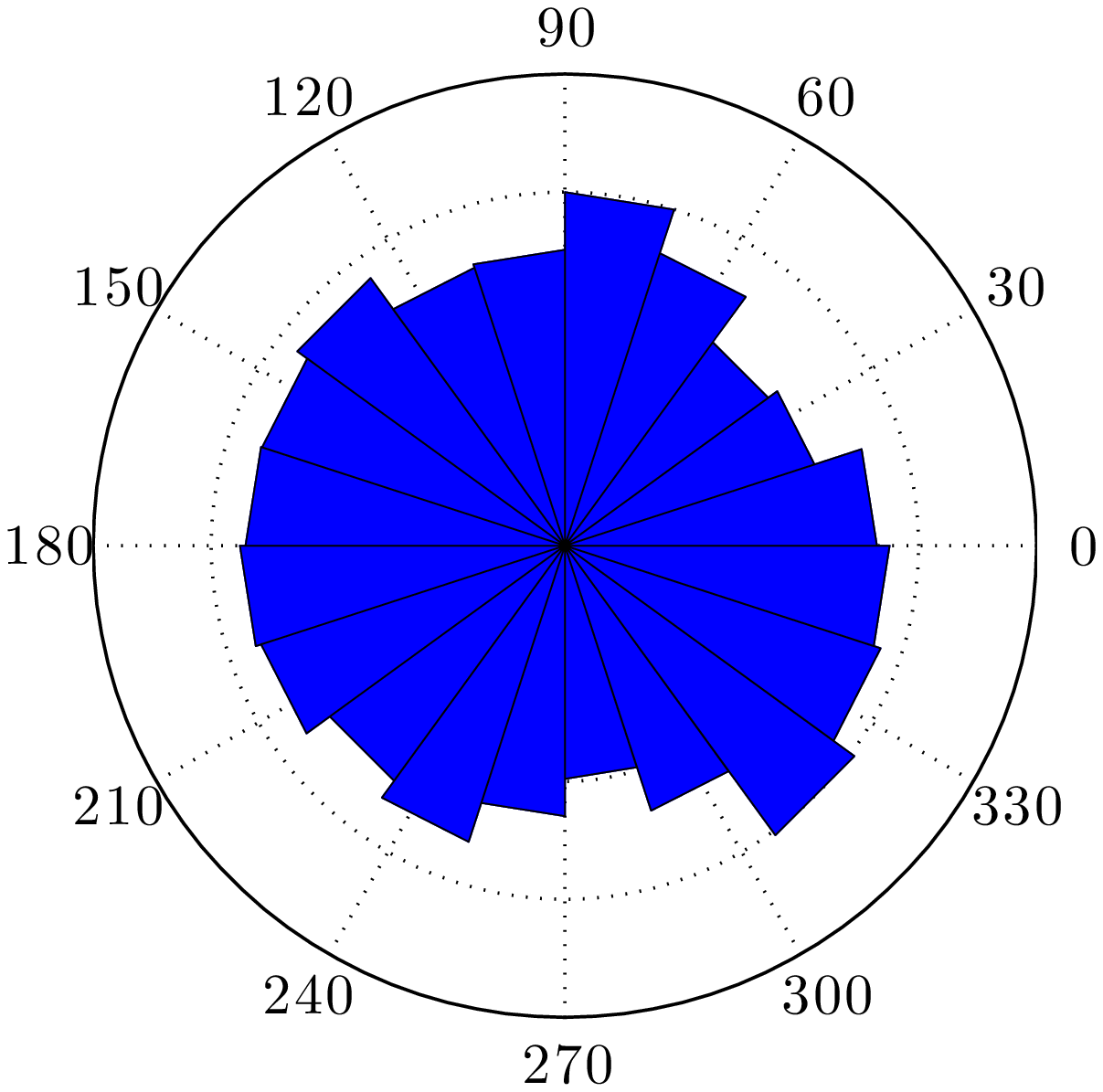}
\includegraphics[width = 0.47\columnwidth]{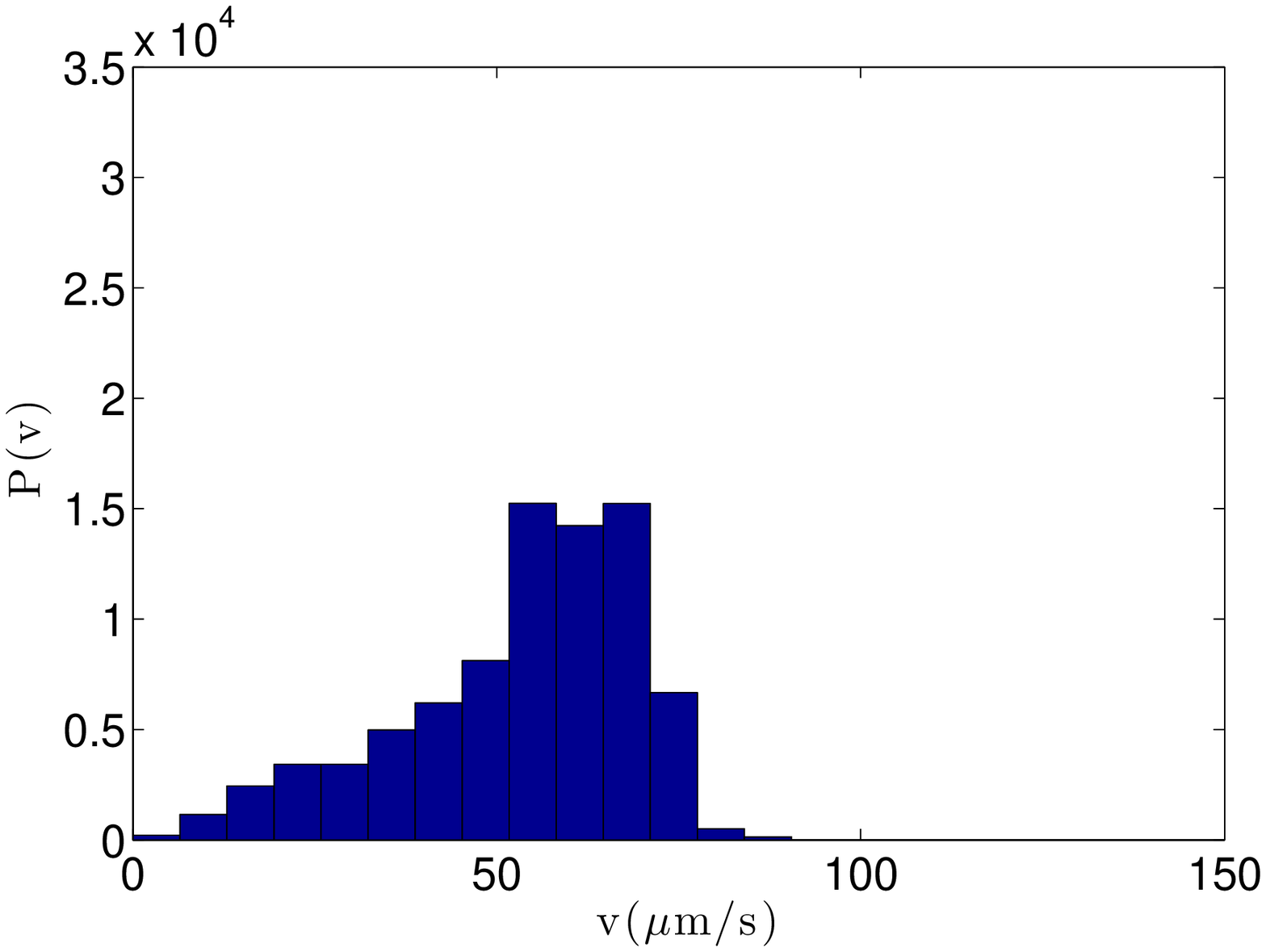}
(c)\includegraphics[width =1.1\columnwidth]{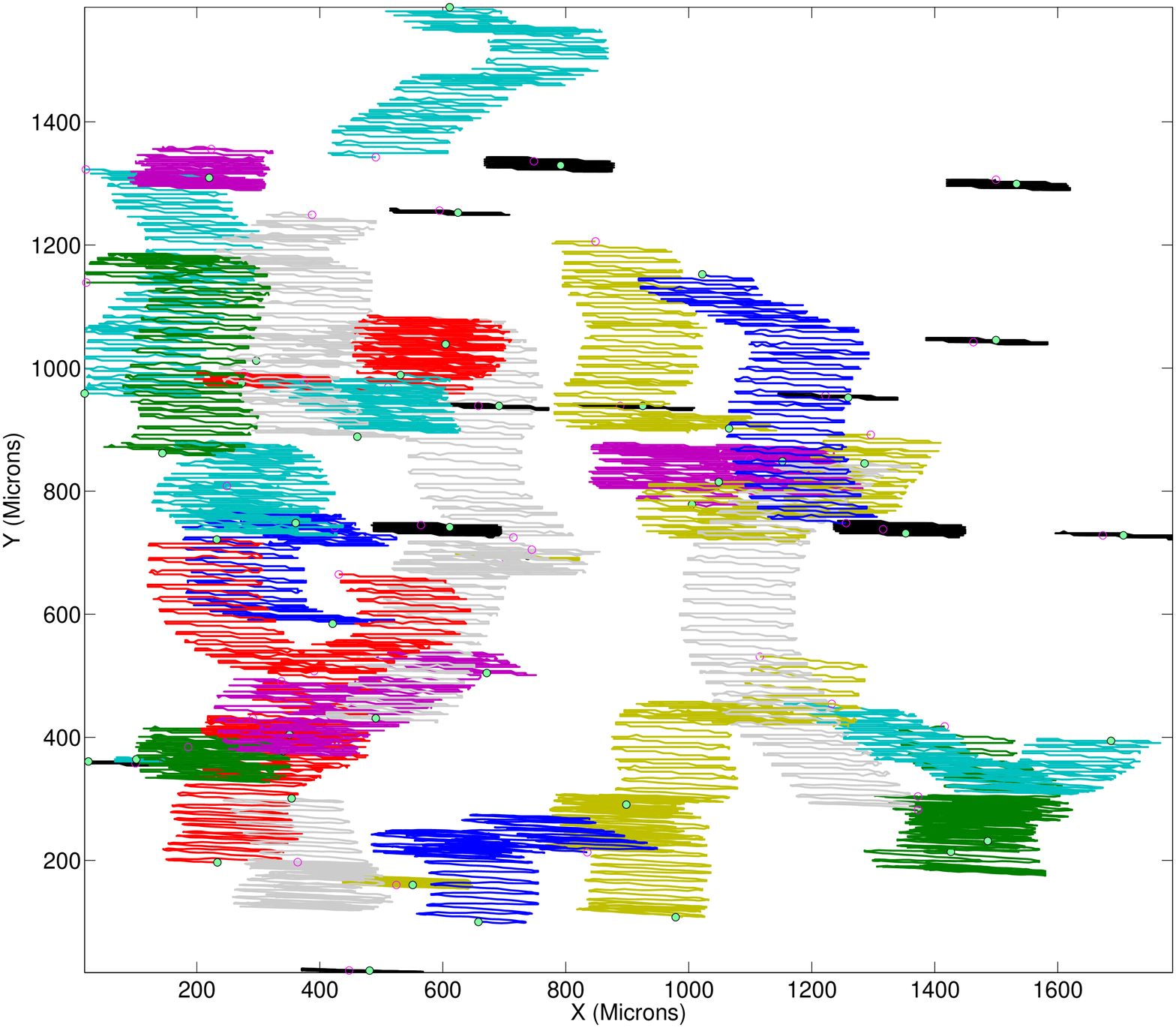}
\caption{\label{Fig1_tracks} Trajectory orientation and speed PDFs before (a) and after (b) oscillatory flow ($A=224$ $\mu$m, $f=2$ Hz as in figure 2c of main text). When oscillatory shear ceases, trajectory isotropy is recovered. (c) Swimmer trajectories (colour) and colloids (black) for $A=150$ $\mu$m, $f=2$ Hz. The latter passively follow the oscillatory flow.}
\end{figure*}

In order to establish that the swimming of {\it Dunaliella salina} in the absence of flow was not affected by exposure to oscillatory shear, the trajectory orientation and swimming statistics of cells were monitored before and after every shearing experiment, see Figure (\ref{Fig1_tracks}). Isotropy and mean speed return to normal once the oscillatory flow is halted. The only effect is a drop in the total number of swimmers in the field of view, probably a result of cell loss in the field of view generated by the anisotropic migration caused by the oscillatory shear (as discussed in the conclusions of the main text). The same (\ref{Fig1_tracks}) shows the simultaneous tracking of swimming algae and PEGylates polystyrene colloids, verifying they passively follow the flow, as expected. 

\section{Simulation model}

\begin{figure}
\includegraphics[width =\columnwidth]{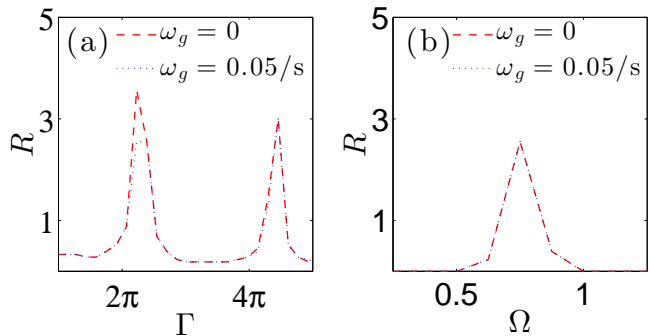}
\caption{\label{Fig2_gravitaxis} Alignment ratio simulations with and without gravitaxis, which makes
little difference to ordering dominated by flow and helical contributions.}
\end{figure}

As mentioned in the main text, mechanisms for the experimental trajectory ordering of {\it Dunaliella salina} in oscillatory shear flows were explored by numerical simulations of helical swimmers. The equations of motion of a swimmer were based on a recently proposed model by Bearon \citep{Bearon13}. In this model (notation as in the main text), cell position $\mathbf{r}$ evolves by swimming with mean speed $v$ along the direction $\mathbf{p}$. Through an intrinsic torque due to nonplanar swimming the cell also generates rotation about a direction $\mathbf{n}$. The dynamics of swimmers in a flow $\mathbf{u}$ is governed by
\begin{eqnarray} \label{eq:helpos}
\dot {\mathbf{r}}&=& \mathbf{u}+ v \mathbf{p}\\
\dot{\mathbf{p}} &=& \boldsymbol\omega_c \times \mathbf{p}\\\label{eq:helorient}
\dot{\mathbf{n}} & = &\boldsymbol\omega_c \times \mathbf{n}\label{eq:heln}
\end{eqnarray}
where the angle between $\mathbf{p}$ and $\mathbf{n}$ is fixed and given by $\mathbf{p}\cdot \mathbf{n}=\cos{\beta}$ and $\boldsymbol\omega_c$ is the net angular velocity of a cell resulting from the balance of torques on it. Bearon considered external torques on the cells due to gravity on bottom heavy spherical swimmers in a flow, so that $\boldsymbol\omega_c=\omega_g \mathbf{p} \times \ez + \omega_h \mathbf{n}+\frac{1}{2}\boldsymbol\omega$, where $\omega_g$ is the gravitational reorientation frequency, $\omega_h$ is the intrinsic angular speed of helical rotation and $\omega$ is the flow vorticity (horizontal and vertical linear shear flows were solved for in \citep{Bearon13}). The model neglects known effects, such as cell shape and stochastic reorientation. These could easily included (e.g. as in \cite{Crozeetal13}), but to capture a minimal mechanism for our observations we neglect them here, and additionally neglect gravitational torques. These torques will only act on cells swimming with a significant vertical component. For these, gravitational reorientation of the cells acts on timescales $1/\omega_g \sim 10$s \cite{CrozeBearonBeesinprep}, which are large compared to the flow timescales for most driving frequencies considered. Thus, flow reorientation is expected to dominate the dynamics, as verified from simulations (see Figure \ref{Fig2_gravitaxis}). 
We then assume the simple oscillatory flow $\mathbf{u}(t)=\gamdinf Z \cos (\omega_d t) \ex$ (vorticity $\boldsymbol\omega(t)=\gamdinf \cos (\omega_d t) \ey$), where $\gamdinf=\omega_d A/H$ is the maximum shear rate for driving angular frequency $\omega_d=2 \pi f_d$ and  amplitude $A$. Nondimensionalising in terms of these time ($1/\omega_d$) and length ($A$) scales, equations (\ref{eq:helpos}-\ref{eq:heln}) reduce to equations (1) and (2) of the main text
\begin{eqnarray}
&&\dot {\mathbf{r}}= \Gamma \cos (t) Z\ex +\nu \mathbf{p}\label{eq:posvect}\\
&&\dot{\mathbf{p}} = \left[\frac{\Gamma}{2}  \cos (t) \ey+\Omega^{-1}\mathbf{n} \right]\times \mathbf{p}\label{eq:orientvectp}\\
&& \dot{\mathbf{n}}  = \frac{\Gamma}{2}  \cos (t)\ey \times \mathbf{n} \label{eq:orientvectn}
\end{eqnarray}
where, as in the main text, we define the dimensionless driving amplitude/shear rate $\Gamma= A/H$ and frequency $\Omega= \omega_d/\omega_h$. The parameter $\nu=v/(A \omega_d)$ gauges the relative magnitude of advection by swimming to advection by flow. Re-writing $\nu=\nu_0/\Omega\Gamma$, with $\nu_0=v/(H \omega_h)$ a constant, we see that $\nu$ cannot be varied independently of the dimensionless driving parameters. In \citep{Bearon13}, where gravity can play a significant role, the ODEs for the components of (\ref{eq:posvect}-\ref{eq:orientvectn}) were derived in a Cartesian coordinate frame where Euler angles are defined with respect to the vertical. We consider instead Cartesian frame with Euler angles $\Theta$ and $\Phi$ describing the helical rotation direction defined from the vorticity direction $\ey$ along the $Y$-axis: $\mathbf{n}=(\sin\Theta_n\sin\Phi_n, \cos\Theta_n, \sin\Theta_n \cos\Phi_n)$, as described in the main text. This choice simplifies the angular equations of motions and makes the dynamics more physically transparent.The equations of motion for the swimmers then become
\begin{eqnarray} 
\dot{X}&=& \nu \sin\Theta_n \sin\Phi_n +  \Gamma \cos (\omega_d t) Z \label{eq:jefftrajX}\\
\dot{Y}&=& \nu \cos\Theta_n  \label{eq:jefftrajY}\\
\dot{Z}&=& \nu \sin\Theta_n \cos\Phi_n \label{eq:jefftrajZ}\\
\dot{\Theta}_n&=&0;\,\,\dot{\Phi}_n=\frac{\Gamma}{2} \cos (t) \label{eq:jefftrajphi}\\
\dot{\Theta}_p&=& \Omega^{-1}\sin\Theta_n \sin(\Phi_n-\Phi_p)\label{eq:heltheta}\\ 
\dot{\Phi}_p&=&\Gamma \cos (t)/2\nonumber\\  \label{eq:helphi}
&+&\Omega^{-1}[\cos\Theta_n-\sin\Theta_n\cot\Theta_p\sin(\Phi_n+\Phi_p)]
\end{eqnarray}
In the limit $\beta\to0$, helical effects are negligible. In this limit, losing subscripts and defining $\nu_\perp=\nu\sin\Theta_0$, $\nu_\parallel=\nu\cos\Theta_0$, equations (\ref{eq:jefftrajX}--\ref{eq:helphi}) reduce to equations $(3-6)$ in the main text, where the dynamics for $\Theta$, $\Phi$ and $Y$ are trivially solved for analytically. The non-integrable equations of system (\ref{eq:jefftrajX}--\ref{eq:helphi}) for swimming cells were solved numerically with MATLAB (Mathworks, Natick, MA, USA) using a Runge-Kutta-Fehlberg (RK45) method in a $1800\times1800\times100\mu$m periodic box. Realistic {\it D. salina} mean swimming parameters (along-helix speed $v=80\mu$m/s and intrinsic frequency $f_d=2$Hz) were used \citep{DDMhelicalinprep}, ignoring phenotypic variations across the population. The helical angle between the swimming direction $\mathbf{p}$ and the helical axis of rotation $\mathbf{n}$ can be estimated from $\beta=\arccos(v/v_p)=0.5$, where $v_p=60\mu$m/s is the mean progressive speed along the helical axis \citep{DDMhelicalinprep}. Cells were set off at $Z=H/2$, with initial values of $\mathbf{p}$ 2D-isotropic in the horizontal plane and $\mathbf{n}$ to be rotated by $\beta$ out of the plane. Realistic driving parameters ($f_d=1-6$Hz and $A=50-3500\mu$m), and a fixed gap width $100\mu$m were used. The ordering ratios $R$ were obtained from the simulation position data, as for experiments.

\providecommand{\noopsort}[1]{}\providecommand{\singleletter}[1]{#1}%

\end{document}